\documentclass[aps,prl,twocolumn,superscriptaddress]{revtex4-2}
\usepackage{graphicx}
\usepackage{xfrac}
\usepackage{amssymb}
\usepackage{amsmath}
\usepackage{verbatim}
\usepackage[unicode=true, bookmarks=true, bookmarksnumbered=false, bookmarksopen=false, breaklinks=false, pdfborder={0 0 1}, backref=false, colorlinks=true]{hyperref}

\usepackage[normalem]{ulem} 
\usepackage{color} 

\begin{document}

	\title{Dispersive Qubit Readout with Intrinsic Resonator Reset}
	
	\author{M.~Jerger}
	\email[E-mail: ]{m.jerger@fz-juelich.de}
	\affiliation{Institute for Functional Quantum Systems (PGI-13), Forschungszentrum J\"ulich, 52425 J\"ulich, Germany}
	\author{F.~Motzoi}
	\affiliation{Institute of Quantum Control (PGI-8), Forschungszentrum J\"ulich, 52425 J\"ulich, Germany}
    \author{Y.~Gao}
	\affiliation{Institute for Functional Quantum Systems (PGI-13), Forschungszentrum J\"ulich, 52425 J\"ulich, Germany}
	\author{C.~Dickel}
	\affiliation{II. Physikalisches Institut, Universit\"at zu K\"oln, Z\"ulpicher Str. 77, 50937 K\"oln,  Germany}
	\author{L.~Buchmann}
	\affiliation{Department of Physics and Astronomy, Aarhus University, 8000 Aarhus, Denmark}
	\author{A.~Bengtsson}
	\affiliation{Microtechnology and Nanoscience, Chalmers University of Technology, SE-412 96 G\"oteborg, Sweden}
    \author{G.~Tancredi}
	\affiliation{Microtechnology and Nanoscience, Chalmers University of Technology, SE-412 96 G\"oteborg, Sweden}
    \author{C.W.~Warren}
	\affiliation{Microtechnology and Nanoscience, Chalmers University of Technology, SE-412 96 G\"oteborg, Sweden}
	\author{J.~Bylander}
	\affiliation{Microtechnology and Nanoscience, Chalmers University of Technology, SE-412 96 G\"oteborg, Sweden}
	\author{D.~DiVincenzo}
	\affiliation{JARA-Institute for Quantum Information (PGI-11), Forschungszentrum J\"ulich, 52425 J\"ulich, Germany}
	\author{R.~Barends}
	\affiliation{Institute for Functional Quantum Systems (PGI-13), Forschungszentrum J\"ulich, 52425 J\"ulich, Germany}
	\author{P. A.~Bushev}
	\email[E-mail: ]{p.bushev@fz-juelich.de}
	\affiliation{Institute for Functional Quantum Systems (PGI-13), Forschungszentrum J\"ulich, 52425 J\"ulich, Germany}
	
	\date{\today}
	
	\begin{abstract}
		
		A key challenge in quantum computing is speeding up measurement and initialization. Here, we experimentally demonstrate a dispersive measurement method for superconducting qubits that simultaneously measures the qubit and returns the readout resonator to its initial state. The approach is based on universal analytical pulses and requires knowledge of the qubit and resonator parameters, but needs no direct optimization of the pulse shape, even when accounting for the nonlinearity of the system. Moreover, the method generalizes to measuring an arbitrary number of modes and states. For the qubit readout, we can drive the resonator to $\sim 10^2$ photons and back to $\sim 10^{-3}$ photons in less than $3 \kappa^{-1}$, while still achieving a $T_1$-limited assignment error below 1\%. We also present universal pulse shapes and experimental results for qutrit readout.
		
	\end{abstract}
	\maketitle
	

    \section{Introduction}

    One of the key challenges for scalable quantum computing is speeding up the repetitive measurement and initialization process required for quantum error correction (QEC) algorithms~\cite{Kitaev1997,Fowler2012,Girvin2023}. Fast measurement will decrease readout error as well as reduce decoherence associated with cycle times in QEC. In superconducting quantum systems, however, the measurement speed is typically orders of magnitude slower than one- and two-qubit gates, which have otherwise reached the $\sim 10$\,ns time scale~\cite{Gambetta2016,Walter2017,Google2020,Manucharyan2021,Pan2022_QEC}. A critical complication is that after measurement, a state-dependent residual intra-cavity field holds up the QEC cycle for a time dramatically exceeding the cavity decay time $\tau_c = \kappa^{-1}$. This residual intra-cavity field shifts the resonance frequency of the qubit via ac-Stark interaction, interferes with subsequent operations, and causes decoherence~\cite{Gambetta2006}. To mitigate these effects, the gate set or circuit compiler has to add extra waiting times to empty the cavity between operations~\cite{Google2021,Gambetta2021,Wallraff2022_QEC,Pan2022_QEC}. 
	
    A valuable resource for QEC algorithms is having a short cavity ring-down time $\tau_c$ for dispersive readout, so that algorithms can resume immediately. There are a couple of methods for the fast depletion of the intra-cavity field. For example, a  fast cavity ($\kappa^{-1}\approx10$~ns) coupled to a Purcell filter allows for a high fidelity readout with 0.25\% assignment error within 88 ns~\cite{Wallraff2022_QEC}. Recently, non-linear Purcell filters have been explored to fix the trade-off between residual intra-cavity photon noise and fidelity~\cite{Nakamura2024}. An optimized cavity-qubit design is used for the multiplexed qubit readout in QEC~\cite{Google2021}. The optimization of tailored multi-step pulse shapes~\cite{McClure2016} or a feed-forward technique with state-dependent depletion pulses may furthermore be used for the fast depletion of the intra-cavity field~\cite{Bultink2016}. However, an approach that natively integrates rapid measurement with the cavity reset, and straightforwardly allows for scaling up the complexity of quantum circuits including linear as well as non-linear elements remains an open challenge. 

    \begin{figure}[ht!]
		\includegraphics[width=1\columnwidth]{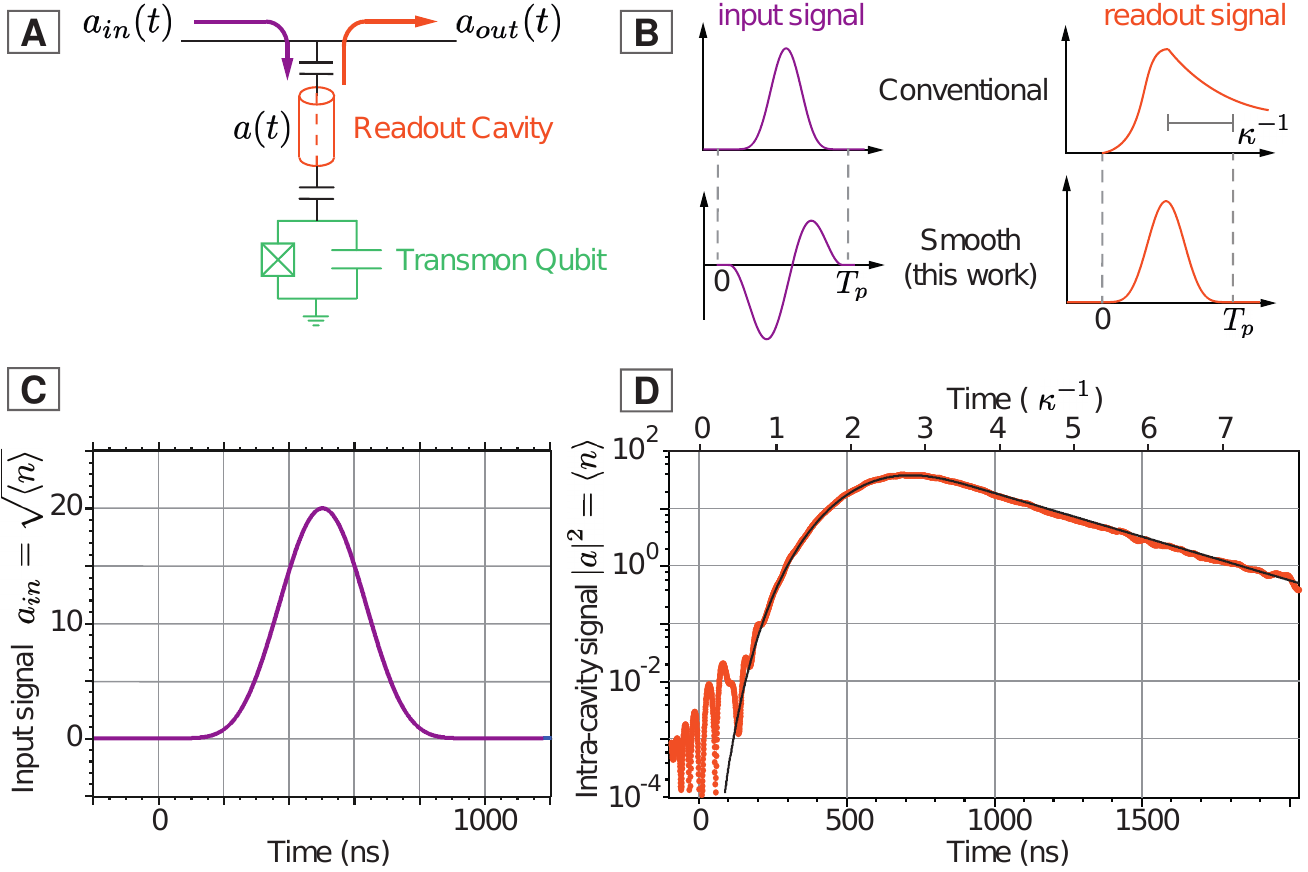}
		\caption {(A) Simplified circuit diagram for the demonstration of dispersive readout. (B) The sketch illustrates a conventional readout with long exponential decay of the intra-cavity field and readout with shaped pulses that reset the resonator to its initial state after the pulse. $T_p$ is the pulse duration. (C and D) The conventional input read-out pulse $a_{in}(t)$ of $T_p=1~$µs causes a long exponential decay of the intra-cavity field $\vert a(t) \vert^2$. The solid line represents a fit of experimental data to a classic response theory, see Eq.(\ref{eq:LTI_td}).}
		\label{Fig:ExpDecay}
	\end{figure}
 
    In this article, we present a simple readout method that combines fast single-shot measurement with state-independent nulling of the readout cavity field using pulses that are analytical. When taking into account the qubit-induced Kerr non-linearity we demonstrate more than a $10^3$-fold residual cavity field reduction immediately after the pulse and assignment error of 0.6\% within a total readout duration of $3\kappa^{-1}$. We also show a three-state readout since the method generalizes to arbitrary numbers of modes and states.

    \section{Experimental demonstration of readout with cavity reset}
    
    \subsection{Transmon device}
    
    We experiment with a conventional circuit QED setup consisting of fixed frequency transmon qubit coupled to a half-wavelength resonator, see the simplified sketch in Fig.~(\ref{Fig:ExpDecay})(A). The resonance frequency of the qubit is measured to be $\omega_{01} / 2\pi = 3947\,$MHz with anharmonicity $\alpha / 2\pi = -232\,$MHz, $T_1 \approx 40$--$75\,\mu$s and $T_2^{\ast}\approx 20$--$40\,\mu$s, varying over time. The resonance frequency of the read-out cavity is $\omega_c / 2\pi = 6041.200$\,MHz and its linewidth is $\kappa /2\pi = 0.5647(3)\,$MHz, corresponding to the time constant of $\tau_c = \kappa^{-1} = 281.2(8)$\,ns. The dispersive shift of the resonator is $\chi/2\pi = 0.299(1)$\,MHz. These numbers are subject to small variations between experimental runs due to the aging of the device. A detailed description of the chip can be found in~\cite{Bengtsson2020} and a description of the experimental setup is presented in Appendix~\ref{app:setup}. The device was fabricated at Myfab Chalmers.

    \subsection{Readout with cavity reset. Concept}

    In the dispersive regime, the Hamiltonian of the qubit-cavity system under excitation of the input field in the rotating frame is well approximated by $\hat{H}_d=\hbar\big( \Delta + \chi \hat{\sigma}_z\big)\hat{a}^+\hat{a}$, where $\Delta = \omega_c-\omega_{01}$ is qubit-cavity detuning. Since the cavity is excited with coherent pulses through the input port, the quantum noise of the microwave signal is additive white vacuum noise which averages to zero, and we can neglect noise operators in our notation. In this simplification, all functions can be considered as the expectation value of coherent optical states, the resonator is described by the classical state-dependent response functions~\cite{BlaisQED2021}. Thus, the relation between input $a_{in}(t)$ and cavity fields $a(t)$ in the time domain is described in terms of linear response theory~\cite{Rytov} by Duhamel's integral~\cite{Duhamel1833}
    \begin{equation}
		a(t) = \int_{-\infty}^{\infty} h(t-\tau) a_{in}(\tau) d\tau,
		\label{eq:LTI_td}
	\end{equation}
	where $h(t)$ is the time-domain response function of the resonator. The Fourier transformation of the Eq.(\ref{eq:LTI_td}) yields the resonator response in frequency space and reads as
	\begin{align}
		a(\omega) &= h(\omega) a_{in}(\omega), \label{eq:LTI_fd} \\
		h(\omega) &= \sqrt{\kappa}/ \left[  \kappa/2 - i(\omega -\omega_i) \right],
    \label{eq:Cavity_Transfer_Omega}	
	\end{align}
	where $\omega_i$ is the state-dependent resonance frequency of the cavity with the boundary condition $a_{out}(\omega) = \sqrt{\kappa} a(\omega)$~\cite{Gardiner1984_in_out,Yurke}. 

    Fig.~\ref{Fig:ExpDecay}(C),(D) demonstrate the response of the cavity to the conventional readout pulse $a_{in}(t)=\sin^{4} \pi t /T_p$ of $T_p=1~\mu$s duration. The process of the initial building of the intra-cavity field $\langle n(t) \rangle = \vert a(t) \vert^2$ is followed by its slow decay with the time constant of $\kappa^{-1}$. It takes $9\kappa^{-1}$ until it fades in the detection noise level corresponding to $\simeq 5 \cdot 10^{-3}$ intra-cavity photons. The dark solid line represents the fit of the experimental data to the response theory described above, where the setup's transmission amplitude, phase/delay, and crosstalk were the only fitting parameters with the known resonator's $\omega_c$ and $\kappa$.

    Following up the shortly outlined theory represented by Eqs.(\ref{eq:LTI_td}),(\ref{eq:LTI_fd}),(\ref{eq:Cavity_Transfer_Omega}), it is possible to tailor an input signal $a_{in}(t)$ that resets the cavity to its initial state at its end ~\cite{Motzoi2018}. For that, we choose the desired cavity response $a=a_T(t)$, referred to as the trial function, that satisfies $a_T(t) = 0\ \forall\ t \leq 0, t \geq T_p$. The input field that generates this cavity response when the qubit is in its ground state can then be calculated by using the inverse Fourier transform:
    \begin{align}
		a_{in}(\omega) &= h^{-1}(\omega)a_T(\omega),\\
		a_{in}(t) &= \bigg[\kappa/2 + i \chi_0 + \frac{d}{dt}\bigg]a_T(t)/\sqrt{\kappa}.
		\label{eq:Smooth_Response1}	
	\end{align}
    The condition that $a_{in}(t)$ be finite-valued and have finite duration $T_p$ implies first-order smoothness of $a_T(t)$ and $\dot{a}_T(0) = \dot{a}_T(T_p) = 0$. Here, we consider all equations in the rotating frame at cavity frequency in the dressed basis. We refer to this technique as Derivative Removal for Annulment of Cavity-Hybridized Measurement Adiabats (\textsc{drachma}).

 	\begin{figure*}[ht!]
		\includegraphics[width=2\columnwidth]{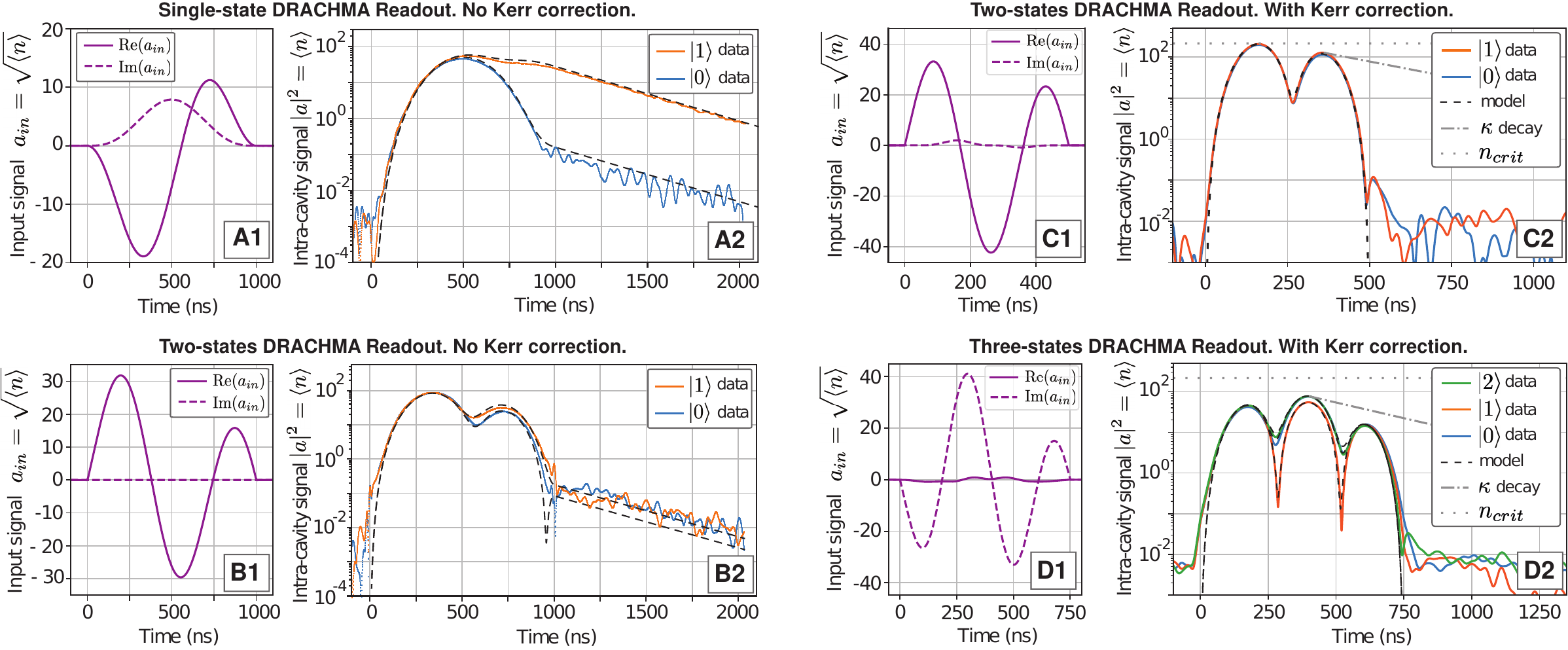}
		\caption { The \textsc{drachma} input fields $a_{in}(t)$ and the dynamics of intra-cavity fields $\langle n (t) \rangle \propto \vert a(t) \vert^2$. (A) Readout with cavity reset for a single state. (B) Readout with cavity reset for the ground and first excited states of the transmon. (C and D) The \textsc{drachma} readout when Kerr-nonlinearity corrections are taken into account. (C) Readout with $>35\,$dB cavity reset for the ground and first excited states of the transmon with $T_p=500\,$ns. (D) Readout with $>30\,$dB cavity reset for the ground, excited and second excited (qutrit) states with $T_p=750\,$ns. Solid lines are experimental data and black dashed lines represent fits to the theory. The dash-dotted line ($\kappa$ decay) shows the conventional exponential decay of the readout cavity from the top of the pulse with decay constant $\kappa^{-1}$. The grey dotted line indicates the field magnitude corresponding to the critical photon number $n_{crit}$.
		}
		\label{Fig:DRACHMA}
	\end{figure*}

    \subsection{Readout with cavity reset. No Kerr correction}

    An example of \textsc{drachma} readout of a single quantum state (ground) is shown in Fig.~\ref{Fig:DRACHMA}(A). The input signal $a_{in}(t)$ of $T_p=1\,\mu$s duration is analytically derived by using Eq.(\ref{eq:Smooth_Response1}) from the trial response function $ a_T(t) = \sin^3 \pi t/T_p$ which satisfies the smoothness condition at the pulse boundaries. Both quadratures of the generated smooth input signal are displayed in the same figure. The input signal is applied at the carrier frequency halfway between the resonator frequencies corresponding to the ground and excited states of the transmon. 
    
    When the transmon is prepared in $\vert 0 \rangle$, see Fig.~\ref{Fig:DRACHMA}(A) the \textsc{drachma} readout causes a 20\,dB partial reset of the intra-cavity field to $\vert a (t) \vert ^2 \simeq 10^{-1}$ photons. Immediately after the pulse, the intra-cavity field experiences the conventional exponential decay. However, when the qubit is prepared in $\vert 1 \rangle$, the readout with \textsc{drachma} pulse works as the ordinary pulse (as plotted in Fig.~\ref{Fig:ExpDecay}). It reaches $10^{-1}$ photons within $6\kappa^{-1}$ after the end of the pulse.

    The above-presented readout technique can be generalized to an arbitrary number of quantum states $N$ in the following way
	\begin{align}
		a_{in}(\omega) &= \bigg [ \prod_{j=0}^{N-1}h^{-1}_{j}(\omega) \bigg] a_T(\omega),\\
		a_{in}(t) &= \bigg [ \prod_{j=0}^{N-1} \bigg(\kappa/2 +i\chi_j + \frac{d}{dt}\bigg)\bigg] a_T(t)/\kappa^{N/2},
		\label{eq:Multiple states}	
	\end{align}
    with $h_{j}(\omega)=\sqrt{\kappa}/ \left[  \kappa/2 - i(\omega -\chi_j) \right]$ being the frequency-domain transfer functions for the particular quantum state $\vert j \rangle $ with dispersive shift $\chi_{j}$ with respect to the carrier frequency. To satisfy the smoothness condition, the trial function $a_T(t)$ and its first $N$ derivatives must all vanish at $t = 0$ and $t = T_p$. An example of such a smooth response function is $ a_T(t) = \sin^m (\pi t/T_p) $, where the exponent $m > N$. 
 
    Fig.~\ref{Fig:DRACHMA}(B) shows the \textsc{drachma} readout of the qubit. The readout pulse $a_{in}(t)$ is analytically calculated from the trial function $a_T(t) = \sin ^3 (\pi t/T_p)$ with experimentally determined state-dependent dispersive shifts $\chi_{0,1}=\pm\chi$ and cavity decay $\kappa^{-1}$. The length of the pulse is set to $T_p=1\,\mu$s. The trajectory of the intra-cavity field for both states follows the optimized pulse shape during the readout pulse duration, resulting at the end of the pulse in the 20\,dB cavity reset to $10^{-1}$ microwave photons. After the end of the readout pulse, the residual intra-cavity field demonstrates conventional exponential decay. The evolution of the intra-cavity fields for the ground and excited states $a_{0,1}(t)$ is very well fitted with the theory where the amplitude and phase/delay of transmission and cross-talk of the setup were the only fitting parameters. 


    \subsection{Readout with cavity reset. With Kerr correction}

    \begin{figure*}[ht!]
        \includegraphics[width=2\columnwidth]{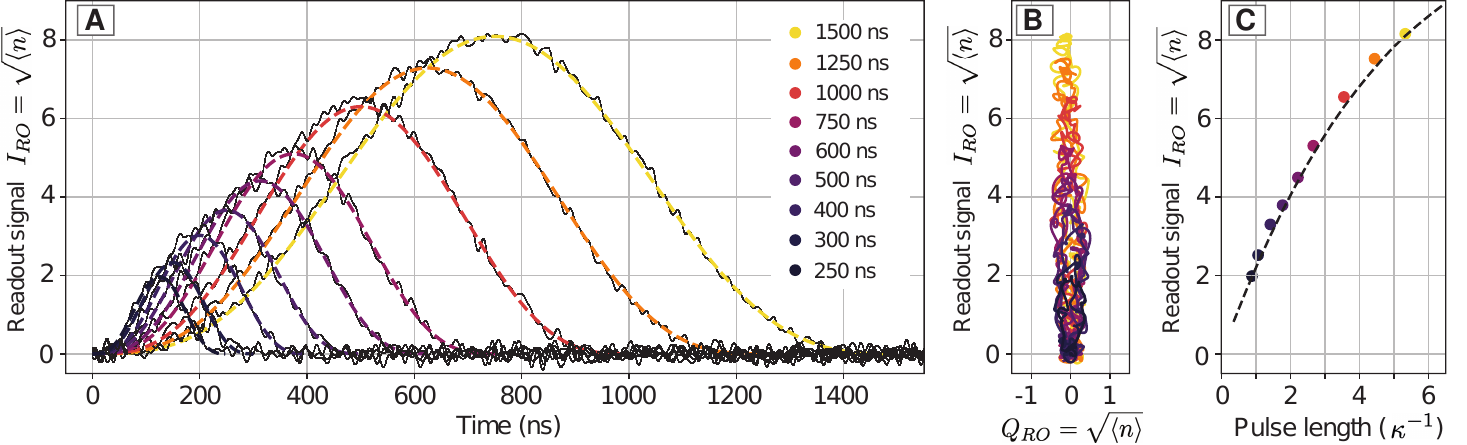}
        \caption {(A) The measured in-phase readout signal $I_{RO} = \operatorname{Re} \big[Z_{RO}(t)\big]$ for the various durations of the \textsc{drachma} pulse between 250 and 1500\,ns. The magnitude of the input signal results in the same number of photons at the maximum of the intra-cavity field. (B) The evolution of the measured readout signal in the $IQ$ plane. (C) The maximum of the readout signal as the function of the pulse duration. The dashed lines show the theoretical predictions from input-output theory. 
        }
        \label{Fig:RO_Signal}
    \end{figure*}
    
    The performance of the above-described readout technique is limited by the additional non-linear shift of the resonator frequency at a relatively strong coherent drive~\cite{Eichler2018}. To achieve the highest measurement contrast with shorter readout pulses, we need to drive the cavity closer to the critical number of photons $n_{c} \simeq 200$. In that case, the system's Hamiltonian needs to include the Kerr non-linear term: $\hat{H}_{tot} = \hat{H}_d+\hbar\zeta \sigma_z (\hat{a}^+)^2\hat{a}^2$, where $\zeta$ is a state-dependent Kerr constant~\cite{Blais2010}. In the semi-classical limit, i.e. in the case of the strong coherent intra-cavity field $a(t)=\langle \hat{a}(t) \rangle \gg 1$, we employ the polaron displacement $\hat{a} \rightarrow \hat{a}+a(t)$. This results in the effective time-dependent ``mean-field" Hamiltonian  
    \begin{equation}
        \hat{H}_\mathrm{eff} = \hbar \Delta \hat{a}^+\hat{a} + \hbar\big(\chi + 4\zeta |a(t)|^2 \big)\hat{\sigma}_z \hat{a}^+\hat{a}, 
        \label{eq:Hamiltonian_Kerr}	  
    \end{equation}
    which contains an additional term describing the power-dependent frequency shift of the resonator. 
    
    We use an iterative approach for the incorporation of the nonlinear shift into the above-described optimized \textsc{drachma} readout. At first, we calculate the actual intra-cavity field of the Kerr-unperturbed resonator at the ground and excited states, by using Eq.(\ref{eq:Multiple states}) for the trial function $a_T(t) = \sin ^3 (\pi t/T_p)$.
    \begin{equation}
        \tilde{a}_{0,1}(t) = \bigg[\kappa/2 +i\chi_{0,1} + \frac{d}{dt}\bigg] a_T(t)/\sqrt{\kappa}.
        \label{eq:intra-cavity_field_qubit_Kerr}	  
    \end{equation}
    For the second iteration, the Kerr-perturbed resonator frequencies at the ground and excited states are used for the calculation of the actual input pulse by using the same Eq.(\ref{eq:Multiple states}):
    \begin{equation}
        a_{in}(t) = \bigg [ \prod_{j=0}^{1} \bigg(\frac{\kappa}{2} +i(\chi_j + 4\zeta_{j} |\tilde{a}_{j}(t)|^2)+ \frac{d}{dt}\bigg)\bigg] a_T(t)/\kappa. 
        \label{eq:Readout_pulse_qubit_Kerr}	
    \end{equation}

    The readout of the transmon qubit with the Kerr-corrected \textsc{drachma} pulse of $T=500\,$ns $(\simeq 2\kappa_c^{-1})$ duration is shown in Fig.~\ref{Fig:DRACHMA}(C). The pulse is rather strong and causes the intra-cavity field to reach the critical photon number $n_{crit}$. The non-linear constants are determined during an optimization routine. At first, both Kerr constants $\zeta_{0,1}$ are set to zero. Then, we perform an incremental scan for both $\zeta$ at a fixed magnitude and duration of the readout pulse while finding the optimal parameter range that results in the maximum depletion of the residual intra-cavity field directly after the pulse. In the presented experiment, the optimal Kerr constants $\zeta_0/2\pi = -175\,$Hz and $\zeta_1/2\pi = -56\,$Hz are comparable to the first-order theoretical estimated value $\tilde{\zeta} = - g^4/\Delta^3 \approx -50\,$Hz. The experimental application of non-linear terms via the \textsc{drachma} method results in the immediate, i.e. faster than $\kappa_c^{-1}$, cavity reset right after the end of the readout pulse even at high photon numbers. The intra-cavity field $a(t)$ demonstrates a significant 35\,dB drop to the noise level of the detection system corresponding to $n\simeq 5 \cdot 10^{-3}$ in the cavity.

    \begin{figure*}[ht!]
        \includegraphics[width=2\columnwidth]{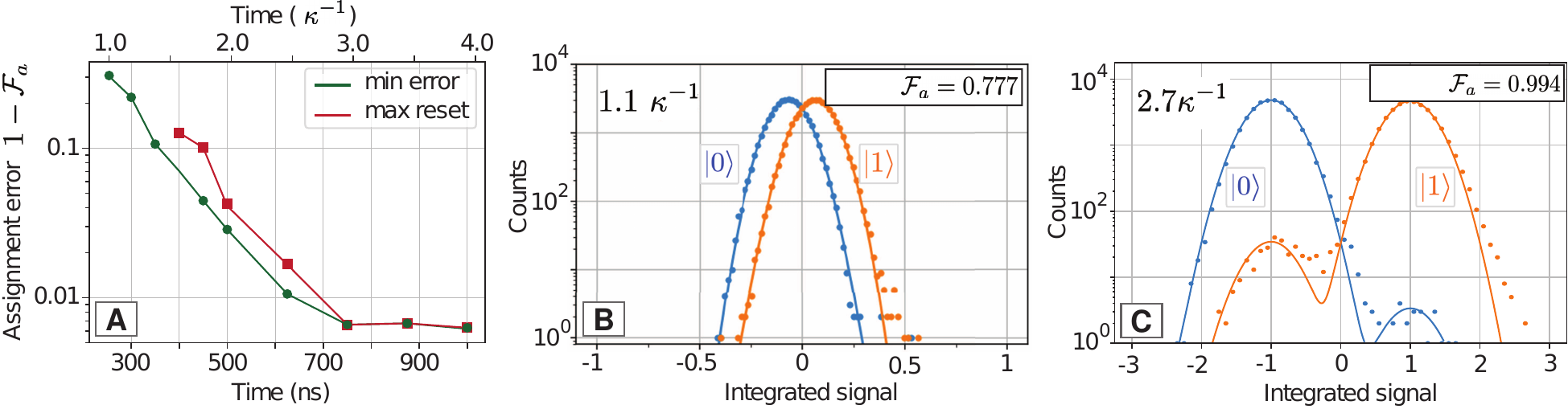}
        \caption {(A) The state assignment error $1-\mathcal{F}_a$ as a function of the \textsc{drachma} pulse length for two cases: (green circles) pulse shapes minimize the assignment error, (red squares) pulse shapes maximize ($>35\,$dB) the reset contrast. (B) and (C)  Experimentally measured histogram of the integrated readout quadrature signal for the selected measurement pulse lengths of $T_p=1.1\kappa^{-1},2.7\kappa^{-1}$. The single-shot measurements are performed with the readout power close to the critical photon number $\langle n_{crit} \rangle \sim 200$. The use of a longer pulse results in better state distinguishability.}
        \label{Fig:Histogram}
    \end{figure*}

    \subsection{Readout of three states with cavity reset}

    The level structure of our transmon device allows for the demonstration of \textsc{drachma} readout for the case of three states, see Fig.~\ref{Fig:DRACHMA}(D). For that purpose, we initially determine the dispersive shifts $\chi_0/2\pi = 0.598$\,MHz, $\chi_1 = 0$ and $\chi_2/2\pi = -0.497$\,MHz relative to the cavity frequency in the first excited state of the qubit and the remaining Kerr constant $\zeta_2 = 60$\,Hz. We use the trial function $a_T(t) = \sin ^4 (\pi t/T_p)$ with a pulse duration of 750\,ns together with non-linear terms to achieve practically an instant reset of the intra-cavity field, where the intra-cavity field drops by more than 30\,dB magnitude with respect to the conventional exponential decay (dash-dotted line $\kappa-$decay) with time constant $\tau_c$. 

    \subsection{Readout signal optimization}

    Fig.~\ref{Fig:RO_Signal}(A) demonstrates the time evolution of the measured in-phase quadrature of the qubit readout signal when applying Kerr-corrected \textsc{drachma} pulses of various durations from 250 to 1500\,ns. The magnitude of the input pulses is chosen to yield the same number of intra-cavity photons for all pulse durations. The acquired quadrature signals are digitally low-pass filtered with the cut-off frequency of 40\,MHz and averaged over $10^6$ repetitions of the experiment. In the case of qubit measurements, the \textsc{drachma} readout signal is engineered in such a way that it is proportional to the trial function
    
     \begin{equation}
	   Z_{RO}(t) \propto \sqrt{\kappa}\big( a_{1}(t)-a_{0}(t) \big) = 2ia_T(t)\chi/\sqrt{\kappa}.
    \label{eq:RO_Qudrature}	
	\end{equation}
   Eq.~\ref{eq:RO_Qudrature} and Fig.~\ref{Fig:RO_Signal}(B) demonstrate clearly that, for a purely real or purely imaginary $a_T$, the readout signal is mainly contained in one of the quadratures. 
   The technique is therefore compatible with homodyne measurement, which can potentially double the measurement efficiency at the cost of using a single frequency. At very short $T_p$ the growing derivative terms of $a_{in}$ increase the maximum photon number in the cavity but not the rate of information extracted from the qubit, so that the readout signal diminishes faster than linear in $T_p$ if the maximum number of photons in the cavity is held constant, as seen in Fig.~\ref{Fig:RO_Signal}(C). This can be mitigated by increasing either $\kappa$ or $\chi$.

   \subsection{Single-shot performance}
   
   To further quantify the performance of the \textsc{drachma} read-out, we now measure the assignment fidelity. We carry out a series of single-shot measurements of the qubit with the Kerr-corrected pulses of different durations. In these experiments, the acquired readout signals $a_{out}(t)$ are multiplied with the analytically calculated complex-valued weighting function $W(t) = \left[ Z_{RO}(t) \right]^*$ for the specific pulse shape and integrated. We perform 10$^6$ repetitions of the experiment while preparing a desired quantum state for each pulse length. The resulting ground and excited state histograms are simultaneously fitted to the sum of two Gaussians. From this fit, we determine a measurement assignment error $1-\mathcal{F}_a=\frac{1}{2}\big[P(1|0)+P(0|1)\big]$, see~\cite{Gambetta2007}, and plot it as a function of readout pulse duration, see Fig.~\ref{Fig:Histogram}(A). The green circles correspond to the readout which minimizes the assignment error, and red squares indicate the case when the cavity experiences a reset by more than 35\,dB. For both cases, the decrease of the assignment error reflects the increase of the information signal for a longer period of the readout/integration time~\cite{Walter2017}. The assignment error reaches its minimal level of 0.6\% for readout durations between $2.7{\kappa}^{-1}$ and $3.5{\kappa}^{-1}$.
    
   Figures~\ref{Fig:Histogram}(B) and (C) demonstrate the experimentally measured histograms of the integrated readout quadrature for the pulse duration of $1.1{\kappa}^{-1}$ and $2.7{\kappa}^{-1}$. The magnitude of the readout signal is chosen such that the intra-cavity field attains the critical number of photons. The specific signal shape also results in the significant $>30$\,dB suppression of the intra-cavity field. The assignment error experiences some fluctuations on the level of $\simeq 0.2$\% which can be attributed to the fluctuations of $T_1$. We believe, that the experimentally demonstrated high assignment fidelity is still largely limited by the low $\chi$ and low amplifier efficiency (17\%), and increasing these would likely boost the assignment fidelity further.

    \section{Conclusion}
    
    The presented readout method can be easily generalized to an arbitrary number of quantum states and resonators, including linear~\cite{Wallraff2018} and non-linear Purcell filters~\cite{Nakamura2024}. Since it is based on the general system's response theory, it can be easily applied to any other physical realization of a quantum system dispersively coupled to a readout resonator system such as ions~\cite{Keller2022}, atoms~\cite{Wallraff2019_Rb,Kessler2021}, opto-mechanics~\cite{Kippenberg_Review2014}, spins in solids~\cite{Bertet2020}, and semiconductor devices~\cite{Vandersypen_Review2019}. 
    
    In conclusion, we have demonstrated a fast, high-fidelity, dispersive readout of transmon qubit with intrinsic cavity reset. Importantly, our approach for cavity reset uses pulses that are analytical. Moreover, only two essential parameters are needed: the dispersive shifts $\chi_j$ of the cavity and its decay rate $\kappa$. And finally, our method requires no complex optimization~\cite{McClure2016}. The crucial step in our method is the introduction of the qubit-induced non-linearity which allows deployment of short pulses of large amplitudes. The presented methodology provides strong and rapid suppression of the cavity ring-downs, and it is well suited for the direct implementation at scale in QEC~\cite{Google2021,Pan2022_QEC, Wallraff2022_QEC}.
    
    This work is supported by the EU through HORIZON-CL4-2022-QUANTUM-01-SGA Project under Grant 101113946 OpenSuperQPlus100  and HORIZON-CL4-2021-DIGITALEMERGING-02-10 Grant Agreement 101080085 QCFD, by DPG under Germany's Excellence Strategy – Cluster of Excellence Matter and Light for Quantum Computing (ML4Q) EXC 2004/1 – 390534769, by the BMBF through the project 13N15685 (GEQCOS), and within the framework program "Quantum technologies – from basic research to market" (Project QSolid, Grant No. 13N16149). We also acknowledge funding from the Knut and Alice Wallenberg Foundation. 

\appendix

\section{Experimental setup}
\label{app:setup}

\begin{figure}[ht!]
		\includegraphics[width=1\columnwidth]{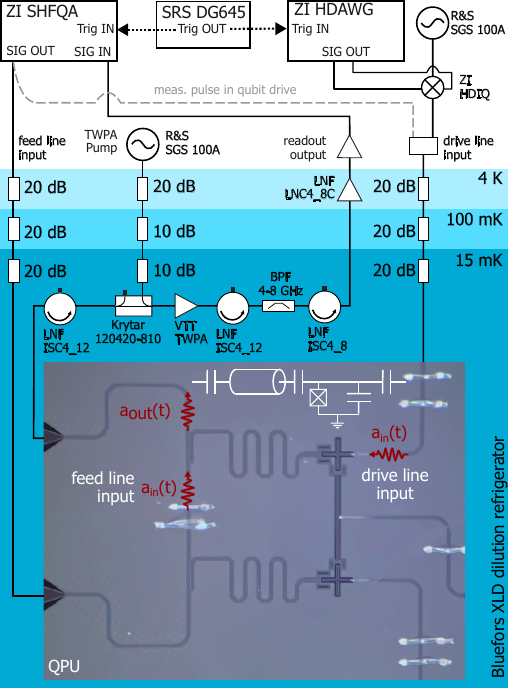}
		\caption {Wiring diagram of the smooth read-out experiment specifying the electronics and components. The readout input pulse  $a_{in}(t)$ is synthesized with the quantum state analyzer (SHFQA) and is applied in two different configurations: through the feedline input or via the qubit drive. The read-out signal $a_{out}(t)$ is amplified and then processed by the SHFQA. The baseband qubit control pulses are generated by HDAWG and upconverted to a carrier frequency. Microscope image of QPU with two transmon qubits. Only the top qubit is used for the experiment. The simplified circuit diagram of the transmon qubit coupled to the resonator (white).}
		\label{Fig:Setup}
	\end{figure}

The experimental setup and a photograph of the superconducting chip are shown in Fig.~\ref{Fig:Setup}. The superconducting device is packaged, protected from spurious magnetic noise by aluminum and cryoperm $\mu$--metal shields and thermally anchored to the mixing chamber of the BlueFors XLD dilution refrigerator. The qubit control drive is synthesized by a Z\"urich Instruments HDAWG signal generator in the baseband, up-converted to the qubit frequency, attenuated, and applied to the charge line of the transmon. The duration of the Gaussian $\pi$-pulse is set to 26.7\,ns. The read-out probing pulse is produced by a Z\"urich Instruments SHFQA and either applied to the common resonator feed line or via the charge line of the qubit~\cite{Mottonen2019}. Both ways result in the same experimental outcome in terms of cavity reset and assignment fidelity, but the latter removes most of $a_{in}$ from the output for visualization purposes. The resulting output signal $a_{out}(t)$ from the common feed line is initially amplified by a traveling wave parametric amplifier (TWPA) with a gain of around 10\,dB. The TWPA is pumped by a continuous tone at 5.3 GHz with a power of about -110\,dBm. After additional amplification at 4\,K by a HEMT (Low Noise Factory LNF-LNC4-8C) and at room temperature the signal and detected by the SHFQA. The cold amplification stages are separated by two cryogenic isolators. The final detection steps include averaging and digital band-pass filtering when time traces are recorded, and weighted integration in case of single-shot measurements. The detection efficiency of the setup is calibrated by using conventional power-spectral density measurements, see~\cite{Walter2017, Bultink2018}, and attains $\eta = 17$\%. 

    \section{Calibration of number of photons}

    \begin{figure}[ht!]
		\includegraphics[width=1\columnwidth]{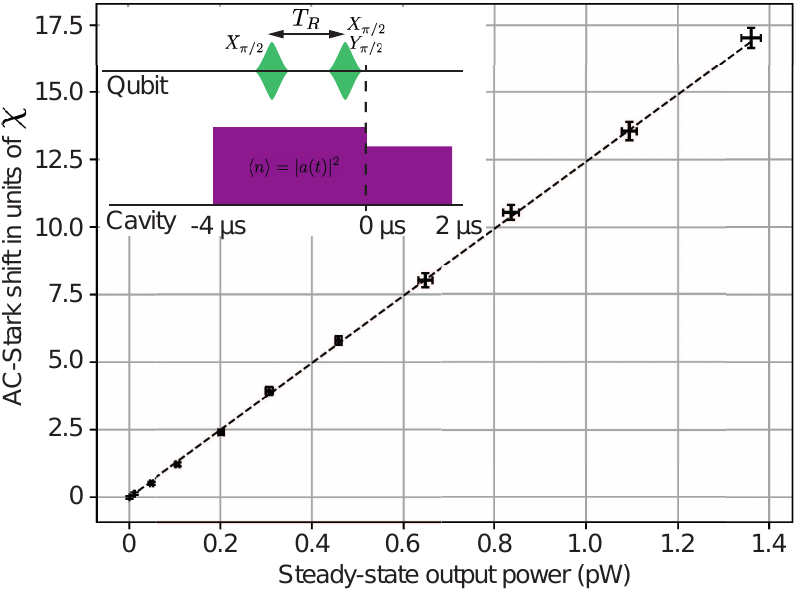}
		\caption {(a) Calibration of the intra-cavity photon number $\langle n(t) \rangle = \vert a(t) \vert^2$ by measuring the AC-Stark shift in a Ramsey experiment versus steady-state output power. Inset: the experimental pulse sequence. A long 4\,µs cavity pulse builds up the steady state population with variable magnitude, which overlaps with a Ramsey sequence at the end of the pulse. The qubit state is measured with a square pulse of 2 µs\,duration.}
		\label{Fig:Cavity-Calibration}
    \end{figure}

    The calibration of the photon number in the cavity requires a two-step procedure. During the first step, we calibrate the total gain of the system, which relates the magnitudes and phases of output and input signals at the SHFQA: $a_{out}(t) = \alpha e^{-i \phi} a_{in}(t) + \beta e^{-i\theta}a(t)$, where $\alpha$, $\beta$ are some coefficients and $\phi$, $\theta$ are the acquired phases. To find $\beta$ and $\theta$ we apply square pulses of variable amplitude and fit the decaying intra-cavity fields after the excitation. The $\alpha$ and $\phi$ coefficients are found by detecting the output signal for a very short 50\,ns pulse, much shorter than $\kappa^{-1}$. In the single-ended driving configuration, i.e. excitation of the cavity through the read-out line, $\beta \approx \alpha/2$ since the driving signal can travel directly to the output and half the signal leaving the cavity travels back to the input, with small corrections due to cross-talk and other loss channels. In the case of driving through the charge line of the qubit $\alpha \ll \beta$.

    During the second step, we calibrate the dependence of the received signal power at the SFHQA input on the number of photons in the cavity. For that, at $t=-4$\,µs the cavity is excited with a 4\,µs long square pulse. This pulse builds up the steady state population of the intra-cavity field in about 1\,µs. Right before the pulse is finished we deploy the standard Ramsey sequence to determine the AC-Stark shift of the qubit $\Delta_{AC}=2\chi \langle n \rangle$~\cite{Blais_AC-Stark2011}. Two $\pi/2$ pulses with separation $T_R = 50-500$\,ns are applied with 10\,MHz detuning from the 01 transition followed by a 2\,µs square measurement pulse for the qubit state measurement. Fig.~\ref{Fig:Cavity-Calibration} shows the linear dependence of the AC-Stark shift on the steady-state output power, which yields 12.4(1) photons per pW of detected power at SHFQA. 

\bibliography{Readout_9}

\begin{thebibliography}{10}

\bibitem{Kitaev1997}
A.~Yu. Kitaev.
\newblock {\em Quantum Error Correction with Imperfect Gates}, pages 181--188.
\newblock Springer US, Boston, MA, 1997.

\bibitem{Fowler2012}
Austin~G. Fowler, Matteo Mariantoni, John~M. Martinis, and Andrew~N. Cleland.
\newblock Surface codes: Towards practical large-scale quantum computation.
\newblock {\em Phys. Rev. A}, 86:032324, Sep 2012.

\bibitem{Girvin2023}
Steven~M. Girvin.
\newblock {Introduction to quantum error correction and fault tolerance}.
\newblock {\em SciPost Phys. Lect. Notes}, page~70, 2023.

\bibitem{Gambetta2016}
Maika Takita, A.~D. C\'orcoles, Easwar Magesan, Baleegh Abdo, Markus Brink,
  Andrew Cross, Jerry~M. Chow, and Jay~M. Gambetta.
\newblock Demonstration of weight-four parity measurements in the surface code
  architecture.
\newblock {\em Phys. Rev. Lett.}, 117:210505, Nov 2016.

\bibitem{Walter2017}
T.~Walter, P.~Kurpiers, S.~Gasparinetti, P.~Magnard,
  A.~Poto\ifmmode~\check{c}\else \v{c}\fi{}nik, Y.~Salath\'e, M.~Pechal,
  M.~Mondal, M.~Oppliger, C.~Eichler, and A.~Wallraff.
\newblock Rapid high-fidelity single-shot dispersive readout of superconducting
  qubits.
\newblock {\em Phys. Rev. Applied}, 7:054020, May 2017.

\bibitem{Google2020}
B.~Foxen, C.~Neill, A.~Dunsworth, P.~Roushan, B.~Chiaro, A.~Megrant, J.~Kelly,
  Zijun Chen, K.~Satzinger, R.~Barends, F.~Arute, K.~Arya, R.~Babbush,
  D.~Bacon, J.~C. Bardin, S.~Boixo, D.~Buell, B.~Burkett, Yu~Chen, R.~Collins,
  E.~Farhi, A.~Fowler, C.~Gidney, M.~Giustina, R.~Graff, M.~Harrigan, T.~Huang,
  S.~V. Isakov, E.~Jeffrey, Z.~Jiang, D.~Kafri, K.~Kechedzhi, P.~Klimov,
  A.~Korotkov, F.~Kostritsa, D.~Landhuis, E.~Lucero, J.~McClean, M.~McEwen,
  X.~Mi, M.~Mohseni, J.~Y. Mutus, O.~Naaman, M.~Neeley, M.~Niu, A.~Petukhov,
  C.~Quintana, N.~Rubin, D.~Sank, V.~Smelyanskiy, A.~Vainsencher, T.~C. White,
  Z.~Yao, P.~Yeh, A.~Zalcman, H.~Neven, and J.~M. Martinis.
\newblock Demonstrating a continuous set of two-qubit gates for near-term
  quantum algorithms.
\newblock {\em Phys. Rev. Lett.}, 125:120504, Sep 2020.

\bibitem{Manucharyan2021}
Quentin Ficheux, Long~B. Nguyen, Aaron Somoroff, Haonan Xiong, Konstantin~N.
  Nesterov, Maxim~G. Vavilov, and Vladimir~E. Manucharyan.
\newblock Fast logic with slow qubits: Microwave-activated controlled-z gate on
  low-frequency fluxoniums.
\newblock {\em Phys. Rev. X}, 11:021026, May 2021.

\bibitem{Pan2022_QEC}
Youwei Zhao, Yangsen Ye, He-Liang Huang, Yiming Zhang, Dachao Wu, Huijie Guan,
  Qingling Zhu, Zuolin Wei, Tan He, Sirui Cao, Fusheng Chen, Tung-Hsun Chung,
  Hui Deng, Daojin Fan, Ming Gong, Cheng Guo, Shaojun Guo, Lianchen Han, Na~Li,
  Shaowei Li, Yuan Li, Futian Liang, Jin Lin, Haoran Qian, Hao Rong, Hong Su,
  Lihua Sun, Shiyu Wang, Yulin Wu, Yu~Xu, Chong Ying, Jiale Yu, Chen Zha, Kaili
  Zhang, Yong-Heng Huo, Chao-Yang Lu, Cheng-Zhi Peng, Xiaobo Zhu, and Jian-Wei
  Pan.
\newblock Realization of an error-correcting surface code with superconducting
  qubits.
\newblock {\em Phys. Rev. Lett.}, 129:030501, Jul 2022.

\bibitem{Gambetta2006}
Jay Gambetta, Alexandre Blais, D.~I. Schuster, A.~Wallraff, L.~Frunzio,
  J.~Majer, M.~H. Devoret, S.~M. Girvin, and R.~J. Schoelkopf.
\newblock Qubit-photon interactions in a cavity: Measurement-induced dephasing
  and number splitting.
\newblock {\em Phys. Rev. A}, 74:042318, Oct 2006.

\bibitem{Google2021}
Zijun Chen, Kevin~J. Satzinger, Juan Atalaya, Alexander~N. Korotkov, Andrew
  Dunsworth, Daniel Sank, Chris Quintana, Matt McEwen, Rami Barends, Paul~V.
  Klimov, Sabrina Hong, Cody Jones, Andre Petukhov, Dvir Kafri, Sean Demura,
  Brian Burkett, Craig Gidney, Austin~G. Fowler, Alexandru Paler, Harald
  Putterman, Igor Aleiner, Frank Arute, Kunal Arya, Ryan Babbush, Joseph~C.
  Bardin, Andreas Bengtsson, Alexandre Bourassa, Michael Broughton, Bob~B.
  Buckley, David~A. Buell, Nicholas Bushnell, Benjamin Chiaro, Roberto Collins,
  William Courtney, Alan~R. Derk, Daniel Eppens, Catherine Erickson, Edward
  Farhi, Brooks Foxen, Marissa Giustina, Ami Greene, Jonathan~A. Gross,
  Matthew~P. Harrigan, Sean~D. Harrington, Jeremy Hilton, Alan Ho, Trent Huang,
  William~J. Huggins, L.~B. Ioffe, Sergei~V. Isakov, Evan Jeffrey, Zhang Jiang,
  Kostyantyn Kechedzhi, Seon Kim, Alexei Kitaev, Fedor Kostritsa, David
  Landhuis, Pavel Laptev, Erik Lucero, Orion Martin, Jarrod~R. McClean, Trevor
  McCourt, Xiao Mi, Kevin~C. Miao, Masoud Mohseni, Shirin Montazeri, Wojciech
  Mruczkiewicz, Josh Mutus, Ofer Naaman, Matthew Neeley, Charles Neill, Michael
  Newman, Murphy~Yuezhen Niu, Thomas~E. O’Brien, Alex Opremcak, Eric Ostby,
  Bálint Pató, Nicholas Redd, Pedram Roushan, Nicholas~C. Rubin, Vladimir
  Shvarts, Doug Strain, Marco Szalay, Matthew~D. Trevithick, Benjamin
  Villalonga, Theodore White, Z.~Jamie Yao, Ping Yeh, Juhwan Yoo, Adam Zalcman,
  Hartmut Neven, Sergio Boixo, Vadim Smelyanskiy, Yu~Chen, Anthony Megrant,
  Julian Kelly, and Google~Quantum AI.
\newblock Exponential suppression of bit or phase errors with cyclic error
  correction.
\newblock {\em Nature}, 595:383--387, 2021.

\bibitem{Gambetta2021}
A.~D. C\'orcoles, Maika Takita, Ken Inoue, Scott Lekuch, Zlatko~K. Minev,
  Jerry~M. Chow, and Jay~M. Gambetta.
\newblock Exploiting dynamic quantum circuits in a quantum algorithm with
  superconducting qubits.
\newblock {\em Phys. Rev. Lett.}, 127:100501, Aug 2021.

\bibitem{Wallraff2022_QEC}
Sebastian Krinner, Nathan Lacroix, Ants Remm, Agustin Di~Paolo, Elie Genois,
  Catherine Leroux, Christoph Hellings, Stefania Lazar, Francois Swiadek,
  Johannes Herrmann, Graham~J. Norris, Christian~Kraglund Andersen, Markus
  Müller, Alexandre Blais, Christopher Eichler, and Andreas Wallraff.
\newblock Realizing repeated quantum error correction in a distance-three
  surface code.
\newblock {\em Nature}, 605:669--674, May 2022.

\bibitem{Nakamura2024}
Yoshiki Sunada, Kenshi Yuki, Zhiling Wang, Takeaki Miyamura, Jesper Ilves,
  Kohei Matsuura, Peter~A. Spring, Shuhei Tamate, Shingo Kono, and Yasunobu
  Nakamura.
\newblock Photon-noise-tolerant dispersive readout of a superconducting qubit
  using a nonlinear purcell filter.
\newblock {\em PRX Quantum}, 5:010307, Jan 2024.

\bibitem{McClure2016}
D.~T. McClure, Hanhee Paik, L.~S. Bishop, M.~Steffen, Jerry~M. Chow, and Jay~M.
  Gambetta.
\newblock Rapid driven reset of a qubit readout resonator.
\newblock {\em Phys. Rev. Applied}, 5:011001, Jan 2016.

\bibitem{Bultink2016}
C.~C. Bultink, M.~A. Rol, T.~E. O'Brien, X.~Fu, B.~C.~S. Dikken, C.~Dickel,
  R.~F.~L. Vermeulen, J.~C. de~Sterke, A.~Bruno, R.~N. Schouten, and
  L.~DiCarlo.
\newblock Active resonator reset in the nonlinear dispersive regime of circuit
  qed.
\newblock {\em Phys. Rev. Applied}, 6:034008, Sep 2016.

\bibitem{Bengtsson2020}
Andreas Bengtsson, Pontus Vikst\aa{}l, Christopher Warren, Marika Svensson, Xiu
  Gu, Anton~Frisk Kockum, Philip Krantz, Christian Kri\ifmmode~\check{z}\else
  \v{z}\fi{}an, Daryoush Shiri, Ida-Maria Svensson, Giovanna Tancredi, G\"oran
  Johansson, Per Delsing, Giulia Ferrini, and Jonas Bylander.
\newblock Improved success probability with greater circuit depth for the
  quantum approximate optimization algorithm.
\newblock {\em Phys. Rev. Applied}, 14:034010, Sep 2020.

\bibitem{BlaisQED2021}
Alexandre Blais, Arne~L. Grimsmo, S.~M. Girvin, and Andreas Wallraff.
\newblock Circuit quantum electrodynamics.
\newblock {\em Rev. Mod. Phys.}, 93:025005, May 2021.

\bibitem{Rytov}
S.M. Rytov, Y.A. Kravtsov, and V.I. Tatarskii.
\newblock {\em Principles of Statistical Radiophysics 2: Correlation Theory of
  Random Processes}.
\newblock Springer-Verlag Berlin Heidelberg, 1988.

\bibitem{Duhamel1833}
J.M.C. Duhamel.
\newblock M\'{e}moir sur les vibrations d'un syst\`{e}me quelconque de points
  materiels.
\newblock {\em Journal de L’Ecole Polytechnique}, 14:1--36, 1833.

\bibitem{Gardiner1984_in_out}
M.~J. Collett and C.~W. Gardiner.
\newblock Squeezing of intracavity and traveling-wave light fields produced in
  parametric amplification.
\newblock {\em Phys. Rev. A}, 30:1386--1391, Sep 1984.

\bibitem{Yurke}
B.~Yurke.
\newblock {\em Input-Output theory}.
\newblock Springer-Verlag Berlin Heidelberg, 2004.

\bibitem{Motzoi2018}
F~Motzoi, L~Bruchmann, and C~Dickel.
\newblock Simple, smooth and fats pulses for dispersive measurements in
  cavities and quantum networks.
\newblock {\em arXive: 1809.04116}, 2018.

\bibitem{Eichler2018}
Johannes Heinsoo, Christian~Kraglund Andersen, Ants Remm, Sebastian Krinner,
  Theodore Walter, Yves Salath\'e, Simone Gasparinetti, Jean-Claude Besse,
  Anton Poto\ifmmode~\check{c}\else \v{c}\fi{}nik, Andreas Wallraff, and
  Christopher Eichler.
\newblock Rapid high-fidelity multiplexed readout of superconducting qubits.
\newblock {\em Phys. Rev. Appl.}, 10:034040, Sep 2018.

\bibitem{Blais2010}
Maxime Boissonneault, J.~M. Gambetta, and Alexandre Blais.
\newblock Improved superconducting qubit readout by qubit-induced
  nonlinearities.
\newblock {\em Phys. Rev. Lett.}, 105:100504, Sep 2010.

\bibitem{Gambetta2007}
Jay Gambetta, W.~A. Braff, A.~Wallraff, S.~M. Girvin, and R.~J. Schoelkopf.
\newblock Protocols for optimal readout of qubits using a continuous quantum
  nondemolition measurement.
\newblock {\em Phys. Rev. A}, 76:012325, Jul 2007.

\bibitem{Wallraff2018}
Johannes Heinsoo, Christian~Kraglund Andersen, Ants Remm, Sebastian Krinner,
  Theodore Walter, Yves Salath\'e, Simone Gasparinetti, Jean-Claude Besse,
  Anton Poto\ifmmode~\check{c}\else \v{c}\fi{}nik, Andreas Wallraff, and
  Christopher Eichler.
\newblock Rapid high-fidelity multiplexed readout of superconducting qubits.
\newblock {\em Phys. Rev. Appl.}, 10:034040, Sep 2018.

\bibitem{Keller2022}
Matthias Keller.
\newblock Cavity-qed with single trapped ions.
\newblock {\em Contemporary Physics}, 63:1--14, 2022.

\bibitem{Wallraff2019_Rb}
S.~Garcia, M.~Stammeier, J.~Deiglmayr, F.~Merkt, and A.~Wallraff.
\newblock Single-shot nondestructive detection of rydberg-atom ensembles by
  transmission measurement of a microwave cavity.
\newblock {\em Phys. Rev. Lett.}, 123:193201, Nov 2019.

\bibitem{Kessler2021}
Hans Ke\ss{}ler, Phatthamon Kongkhambut, Christoph Georges, Ludwig Mathey,
  Jayson~G. Cosme, and Andreas Hemmerich.
\newblock Observation of a dissipative time crystal.
\newblock {\em Phys. Rev. Lett.}, 127:043602, Jul 2021.

\bibitem{Kippenberg_Review2014}
Markus Aspelmeyer, Tobias~J. Kippenberg, and Florian Marquardt.
\newblock Cavity optomechanics.
\newblock {\em Rev. Mod. Phys.}, 86:1391--1452, Dec 2014.

\bibitem{Bertet2020}
V.~Ranjan, J.~O'Sullivan, E.~Albertinale, B.~Albanese, T.~Chaneli\`ere,
  T.~Schenkel, D.~Vion, D.~Esteve, E.~Flurin, J.~J.~L. Morton, and P.~Bertet.
\newblock Multimode storage of quantum microwave fields in electron spins over
  100 ms.
\newblock {\em Phys. Rev. Lett.}, 125:210505, Nov 2020.

\bibitem{Vandersypen_Review2019}
Lieven M.~K. Vandersypen and Mark~A. Eriksson.
\newblock Quantum computing with semiconductor spins.
\newblock {\em Physics Today}, 72:38--45, 08 2019.

\bibitem{Mottonen2019}
Joni Ikonen, Jan Goetz, Jesper Ilves, Aarne Ker\"anen, Andras~M. Gunyho, Matti
  Partanen, Kuan~Y. Tan, Dibyendu Hazra, Leif Gr\"onberg, Visa Vesterinen,
  Slawomir Simbierowicz, Juha Hassel, and Mikko M\"ott\"onen.
\newblock Qubit measurement by multichannel driving.
\newblock {\em Phys. Rev. Lett.}, 122:080503, Feb 2019.

\bibitem{Bultink2018}
C.~C. Bultink, B.~Tarasinski, N.~Haandbæk, S.~Poletto, N.~Haider, D.~J.
  Michalak, A.~Bruno, and L.~DiCarlo.
\newblock General method for extracting the quantum efficiency of dispersive
  qubit readout in circuit qed.
\newblock {\em Applied Physics Letters}, 112(9):092601, 2018.

\bibitem{Blais_AC-Stark2011}
F.~R. Ong, M.~Boissonneault, F.~Mallet, A.~Palacios-Laloy, A.~Dewes, A.~C.
  Doherty, A.~Blais, P.~Bertet, D.~Vion, and D.~Esteve.
\newblock Circuit qed with a nonlinear resonator: ac-stark shift and dephasing.
\newblock {\em Phys. Rev. Lett.}, 106:167002, Apr 2011.

\end{thebibliography}
\end{document}